\documentstyle[12pt]{article}
\begin{document}
\begin{center}
{\bf ON THE SPECTRUM OF THE MANY-BODY PAULI PROJECTOR  }\\
\vspace*{0.5cm} E.M. Tursunov\\
\end{center}
 \begin{center}
  Institute of Nuclear Physics,100214, Ulugbek, Tashkent, Uzbekistan
 \end{center}
 \begin{abstract}
  Spectrum of the Pauli projector of a quantum many-body system is
  studied. It is proven that the kern of the complete many-body
  projector is identical to the kern of the sum of two-body
  projectors. Since the kern of the many-body Pauli projector defines an
  allowed subspace of the complete Hilbert space, it is argued that a truncation of the
  many-body model space following the two-body Pauli projectors is a natural way when solving the Schr\"{o}dinger equation for the many-body system. These relations clarify a role of the
  many-body Pauli forces in a multicluster system.
 \end{abstract}
\vskip 5mm

\vskip 5mm {\small {\bf Key words: Pauli projector, many-body quantum system,
forbidden states, allowed subspace}
\vskip 5mm

 Recent  prediction of a quantum phase transition (QPT)  in the $^{12}$C ground state nucleus within the frame of {\it ab initio}  technics 
\cite{serdar16} based on the chiral effective field theory potentials, inspired new research interest on the structure of this important quantum object.  
A question, whether it is possible to observe an affect of the QPT within the framework of the 3$\alpha$ cluster model, is of great importance. 
Another interesting property of this nucleus is its special structure, associated with the Bose-Einstein condensation \cite{review12C}. 
Although the  3$\alpha$ cluster model for the structure of the lowest $^{12}$C states seems very natural due to strong binding of nucleons inside the  $\alpha$ clusters, there are serious problems, associated with the removal of Pauli forbidden states in macroscopic cluster models.   
  
The problem of  removal of Pauli forbidden states (FS) in a quantum 
many body system has been discussed extensively during the long period
\cite{smir74,hor75,schmid82,tur01,fuji06,lash09}. The most
popular system for the study of projection technics is  $^{12}$C  as  3$\alpha$. When using a
deep $\alpha\alpha$- potential of the BFW form \cite{bfw}, there are
three Pauli forbidden states in partial waves $|0S>$
, $|2S>$ and $|2D>$  in the each of two-body $\alpha \alpha$- subsystems. For the realistic description of the system one
has to eliminate all FS from the solution of the three-body
Schr\"{o}dinger equation by using the supersymmetric transformation (SUSY) \cite{baye87}, the
orthogonal condition model (OCM) \cite{sai69} or the method of orthogonalising pseudopotentials (OPP) \cite{kuk78}. The resulting solution of the Schr\"{o}dinger equation
strongly depends on the orthogonalisation method. The first evidence of a strong influence of the orthogonalisation technics on the quality of the three-body wave function was found in the beta decay study of the $^6$He halo nucleus into the $\alpha +d$ continuum \cite{tur06,tur06a}, where the $^6$He nucleus was described as an $\alpha+n+n$ two neutron halo state. Very recently it was found that  the same effect can be observed in the study of the astrophysical capture reaction 
$\alpha+d \to ^6$Li $+\gamma$ within the three-body model \cite{tur16,tur18,tur19}. In both processes the three-body wave function obtained using the SUSY orthogonalisation technics failed to describe the experimental data, while the OPP method yields a very good description. A success of the OPP method is connected with its property to yield a nodal behavior of the three-body scattering and bound state wave functions at short distances due to Pauli blocking, while the SUSY transformation of the initial potential does not keep this realistic property.    

Coming back to the 3$\alpha$ problem, we note that  within the OPP method it was found \cite{tur01}
 that the energy spectrum of the
ground $0^+$ and first excited $2^+_1$ states is
strongly sensitive to the description of the $\alpha \alpha$- Pauli forbidden states.
The alternative direct orthogonalisation method  \cite{fuji06} is based on the 
separation of the complete Hilbert functional space  into the two parts:
allowed and forbidden by the Pauli principle 3$\alpha$ states. 
The allowed subspace is defined by the kern of the complete three-body projector.
However, it was found that in the 3$\alpha$  system there are so called "almost forbidden states"
\cite{fuji06}, which correspond to the almost zero eigen values of the three-body projector,
 $ \hat{P} =\sum_{i}\hat{P}_i$, where
each $\hat{P}_i$  is the projector on Pauli-forbidden states in the i-th
two-body subsystem. A serious problem was to answer the question: to
eliminate these states or to keep them in the three-body model space?  In
the first case one has a strong underbinding , while the second way
results in a large overbinding. An original  solution was suggested
in Ref.\cite{fuji06} to use the microscopic description of the
forbidden states and not to use the FS of the BFW potential. Such a
way gives "normal" three-body FS contrary to the three-body FS
derived from the potential. However, from physical wievpoint, this way is not realistic since the forbidden states should be associated with two-body potentials, which describe the experimental data, binding energies and phase shifts. 

 On the other hand, the complete projector is more than the sum of two-body projectors
 (see Ref. \cite{smir74}):
\begin{equation}
\hat{\Gamma}=\sum_{i=1}^{3} \hat{P}_i- \sum_{i\ne j=1}^{3}
\hat{P}_i\hat{P}_j+ \sum_{i \ne j \ne k=1}^{3} \hat{P}_i\hat{P}_j
\hat{P}_k- \cdots,
\end{equation}
where
\begin{equation}
\hat{P}_i=\sum_{f} \hat{\Gamma}_i^{(f)},
\end{equation}
and $\hat{\Gamma}_i^{(f)}$ is the projecting operator to the
$f$-wave forbidden state in the two-body subsystem $(j+k)$,
$(i,j,k)=(1,2,3)$, and their cyclic permutations. 
 Here two-body projectors do not commute each with other:
 $ \hat{P}_i\hat{P}_j\ne \hat{P}_j \hat{P}_i$    and $\hat{P}_i^ 2= \hat{P}_i$.
  However, they commute  with the complete projector:
  $\hat{P}_i \hat{\Gamma} = \hat{\Gamma} \hat{P}_i = \hat{P}_i$.
The sums on the right hand side of the last equation contain terms like $\hat{P}_1 \hat{P}_2 \hat{P}_1$ due to above noncommutativity.

   One has to note that the method of OPP, as well as the direct diagonalization technics \cite{fuji06}  use
only the first term of the expansion for the operator $\hat{\Gamma}$. A question is, whether the
neglecting of  next terms of the complete projector in these methods
is a good approximation? In other words, are the three-body
Pauli forces negligible? Our estimation for the overlap of the
$|0S>$  forbidden states  from different   subsystems was around
1.367 which means that the terms like $\hat{P}_i \hat{P}_j$  can
give additional non negligible contribution to the projector if they don't mutually cancel each other.
On the other hand, for the 3$\alpha$ system the microscopic calculations show
negligible contribution from three-body Pauli forces $\cite{fuji06}$.
Thus, a possible contribution of the three-body Pauli forces  to the full projector
must be examined in a strong mathematical way.

   A way to relate the spectrum of the complete projector  $\hat{\Gamma}$
   with the sum of the two-body projectors $\hat{P}$ is based on the algebra of the operators
$\hat{P}_i$. A final result can be formulated as a

    { \bf{THEOREM 1}}: The complete many-body projector $\hat{\Gamma}$
    is related to the sum of the two-body projectors $\hat P=\sum_{i}{\hat {P_i}}$ as
   \begin{equation}
       \hat{\Gamma} = 1- \lim_{m \to \infty} (1-\hat{P})^m
   \end{equation}
     {\bf Proof:} We define the operator $\hat{\Gamma}_n$ as a sum
     of the first $n$ terms in the expansion of Eq.(1):
    \begin{eqnarray}
\nonumber
 \hat{\Gamma}_n=\sum_{i=1}^{3} \hat{P}_i- \sum_{i\ne
j=1}^{3} \hat{P}_i\hat{P}_j+ \sum_{i \ne j \ne k=1}^{3}
\hat{P}_i\hat{P}_j
\hat{P}_k- \cdots +    \\
\nonumber
 (-1)^{(n-1)}\sum_{i1\ne i2 \cdots}
\hat{P}_{i1}\hat{P}_{i2}\cdots \hat{P}_{in}
\end{eqnarray}
With this definition, we will prove the relation
 \begin{equation}
  \hat{\Gamma}_m=1- (1-\hat{P})^m
\end{equation}
   for any value of $m$.
       The proof will be done by using the mathematical induction.
First we note that the Eq. (4) is correct for $m=1$. Now we assume
that it is correct for $m=n$ and we prove it for the case $m=n+1$.
By multiplying the operator $\hat{\Gamma}_n$ from the left side by
the two-body operator $\hat{P}$  and using the commutation relations
of the projectors $\hat{P}_i$ we can write the relation:
\begin{eqnarray}
\nonumber
 \hat{P} \hat{\Gamma}_n= \sum_{i=1}^{3} \hat{P}_i
+\sum_{i\ne j=1}^{3} \hat{P}_i\hat{P}_j - \sum_{i\ne j=1}^{3}
\hat{P}_i\hat{P}_j -
 \sum_{i \ne j \ne k=1}^{3} \hat{P}_i\hat{P}_j
\hat{P}_k +
 \sum_{i
\ne j \ne k=1}^{3} \hat{P}_i\hat{P}_j \hat{P}_k+ \cdots +         \\
(-1)^{(n-1)}\sum_{i1\ne i2 \cdots} \hat{P}_{i1}\hat{P}_{i2}\cdots
\hat{P}_{in} \hat{P}_{in+1} =
 \hat{P}+(-1)^{(n+1)}\sum_{i1\ne i2 \cdots}
\hat{P}_{i1}\hat{P}_{i2}\cdots \hat{P}_{in} \hat{P}_{in+1}
\end{eqnarray}
 In the last equation all the terms are canceled except the first and the last
 ones. It gives us the relation:
 \begin{equation}
  \sum_{i1\ne i2 \cdots}
\hat{P}_{i1}\hat{P}_{i2}\cdots \hat{P}_{in} \hat{P}_{in+1} =
(-1)^{(n+1)} \hat{P}(\hat{\Gamma}_n-1)
 \end{equation}

 On the other hand, from the definition of the operator
 $\hat{\Gamma}_n$ one can write:
 \begin{equation}
  \hat{\Gamma}_{n+1}= \hat{\Gamma}_n +
 (-1)^{n}\sum_{i1\ne i2 \cdots}
\hat{P}_{i1}\hat{P}_{i2}\cdots \hat{P}_{in} \hat{P}_{in+1}
 \end{equation}
Now on the basis of the relation (6),  one can write a reccurent formula:
 \begin{equation}
  \hat{\Gamma}_{n+1}-1= (1-\hat{P}) (\hat{\Gamma}_n -1)=(1-\hat{P})^2 (\hat{\Gamma}_{n-1} -1)=\cdots =
(1-\hat{P})^n (\hat{\Gamma}_1 -1).
  \end{equation}
By using the definition of the operator $\hat{\Gamma}_1=\hat{P}$, finallly we come to the relation
\begin{equation}
\hat{\Gamma}_{n+1}=1-(1-\hat{P})^{(n+1)}.
\end{equation}
         The proven relation (3) enables us a way to define the allowed many-body model
space, which corresponds to the kern of the operator $\hat{\Gamma}$.
Thus we come to the

   { \bf{THEOREM 2}:   The kern of the operator $\hat{P}=\sum_{i}{\hat{P}_i}$ is identical to the
    kern of the complete many-body Pauli projector $\hat{\Gamma}$ }.

   {\bf Proof:} Let $\Psi$ belongs to the kern of the operator
   $\hat{P}$, i.e. it is an eigen-function of the operator
   $\hat{P}$ corresponding to the eigen-value $\lambda=0$:
   $\hat{P}\Psi=0$. Then from the relation (3) one can find
   $\hat{\Gamma}\Psi=0$. 


Now, let $\Psi$ belongs to the
 kern of the operator $\hat{\Gamma}$.
It means that there is  a large number $N$ and for any $n>N$:
\begin{equation}   
(1-\hat{P})^n\Psi=\Psi. 
\end{equation}
Multiplying the last equation from both sides by the operator $1-\hat{P}$ one can write:
\begin{equation}
(1-\hat{P})^{n+1}\Psi=(1-\hat{P})\Psi. 
\end{equation}
As well as the condition $n+1>N$ is valid, then
\begin{equation}   
(1-\hat{P})^{n+1}\Psi=\Psi.
\end{equation}   

Now from the last two equations  one can obtain
\begin{equation}
(1-\hat{P})\Psi=\Psi. 
\end{equation}
and consequently, $\hat{P}\Psi=0$. In other words, the function $\Psi$ belongs to the kern of the operator $\hat{P}$.

 This theorem clarifies a role of many-body Pauli forces in multicluster systems.
 Since the kern of the projecting operator defines an allowed subspace, it is good enough to
 expand a probe wave function of the many-body Hamiltonian over the eigen states of the operator
$\hat{P}=\sum_{i}{\hat{P}_i}$ corresponding to the zero eigen-value.
The obtained results indicate that the truncation of the allowed
model space following the operator $\hat{P}$ is a natural procedure.
This way is valid even for the case when the
operators $\hat{P}_i$ and $\hat{P}_j$ overlap strongly.

    \end{document}